\documentclass[reprint,superscriptaddress,amsmath,amssymb,aps]{revtex4-2}
\usepackage[scaled]{beramono}
\usepackage[T1]{fontenc}
\usepackage{graphicx, amsfonts, amsmath, amssymb, cancel, bm}
\usepackage[colorlinks=true]{hyperref}
\usepackage{siunitx}
\usepackage{xcolor}
\usepackage{colortbl}
\usepackage{commath}
\usepackage{bbm}
\patchcmd{\subsubsection}{\itshape}{\bfseries}{}{}

\usepackage[capitalize]{cleveref}

\begin{document}

\title{Static wetting of a barrel-shaped droplet on a soft-layer-coated fiber}

\author{Bo Xue Zheng}
\affiliation{Department of Mathematics, Mechanics Division, University of Oslo, N-0851 Oslo, Norway}

\author{Christian Pedersen}
\affiliation{Department of Mathematics, Mechanics Division, University of Oslo, N-0851 Oslo, Norway}

\author{Andreas Carlson}
\affiliation{Department of Mathematics, Mechanics Division, University of Oslo, N-0851 Oslo, Norway}

\author{Tak Shing Chan}
\email{taksc@uio.no}
\affiliation{Department of Mathematics, Mechanics Division, University of Oslo, N-0851 Oslo, Norway}

\date{\today}

\begin{abstract}
A droplet can deform a soft substrate due to capillary forces when they are in contact. We study the static deformation of a soft solid layer coated on a rigid cylindrical fiber when an axisymmetric barrel-shaped droplet is embracing it. We find that the elastic deformation increases with decreasing rigid fiber radius. Significant disparities of deformation between the solid-liquid side and the solid-gas side are found  when their solid surface tensions are different. When the coated layer is soft enough and the rigid fiber radius is less than the thickness of the coated layer, pronounced displacement oscillations are observed. Such slow decay of deformation with distances from the contact line position suggests a possible long-range interaction between droplets on a soft-layer-coated fiber.
\end{abstract}

\maketitle

\section{Introduction}
A layer of a soft solid material such as gels and elastomers coated on rigid solid substrates,  can be significantly deformed by capillary forces, when a droplet is in contact with it \cite{Park2014,Andreotti2020}. The shape of the deformation has been explored extensively in the last decades, however studies mainly focus on situations of a droplet on a planar substrate \cite{Lester1961,Rusanov1975,Pericet-Camara2008,Yu2009,Pericet-Camara2009,Jerison2011,Das2011,Limat2012,Yu2012,Style2012,Style2013,Lubbers2014,Park2014,Hui2014,Bostwick2014,Dervaux2015,Style2017,Style2018,Masurel2019,Andreotti2020,Choi2021,Yang2021,Chan2022}. Although there have been numerous studies of droplets on rigid fibers \cite{lorenceau2004,Er2013,Chan2020a,Chan2021,Fournier2021,Xu2022}, wetting on a rigid fiber coated with a soft layer has been far less investigated \cite{Guan2020}.  How this geometry modifies the deformation of the coated elastic layer remains unclear, which is the focus of this study.

The wetting of droplets on fibers are ubiquitous in both industrial and natural phenomena,  examples range from the modulation of the mechanical properties of spider silk \cite{Vollrath89},  to the droplet transport on spines of cacti \cite{Liu2015},  to water collection by fog net.  
Compared to a planar surface ,  the slender geometry of fibers modifies the droplet shape \cite{Er2013,Xu2022},  and could induce directional motion of droplets \cite{CARROLL1976488,lorenceau2004,Chan2020,Lee2022}. When it comes to droplet wetting on soft fibers,  most studies have focused on the bending and buckling of flexible fibers which result in, for instance,  the winding around a droplet  \cite{Schulman2017},  the coiling inside a droplet  \cite{Elettro2015} or the modification of droplet morphology by multiple fibers \cite{Duprat2012}.  On the other hand,  a soft fiber with a rigid core, i.e.  a soft-layer-coated fiber, can hardly bend. Hence the interaction of droplets with a soft-layer-coated fiber is expected to be different from that of a totally soft fiber.

\begin{figure}
\begin{center}
\includegraphics[width=0.49\textwidth]{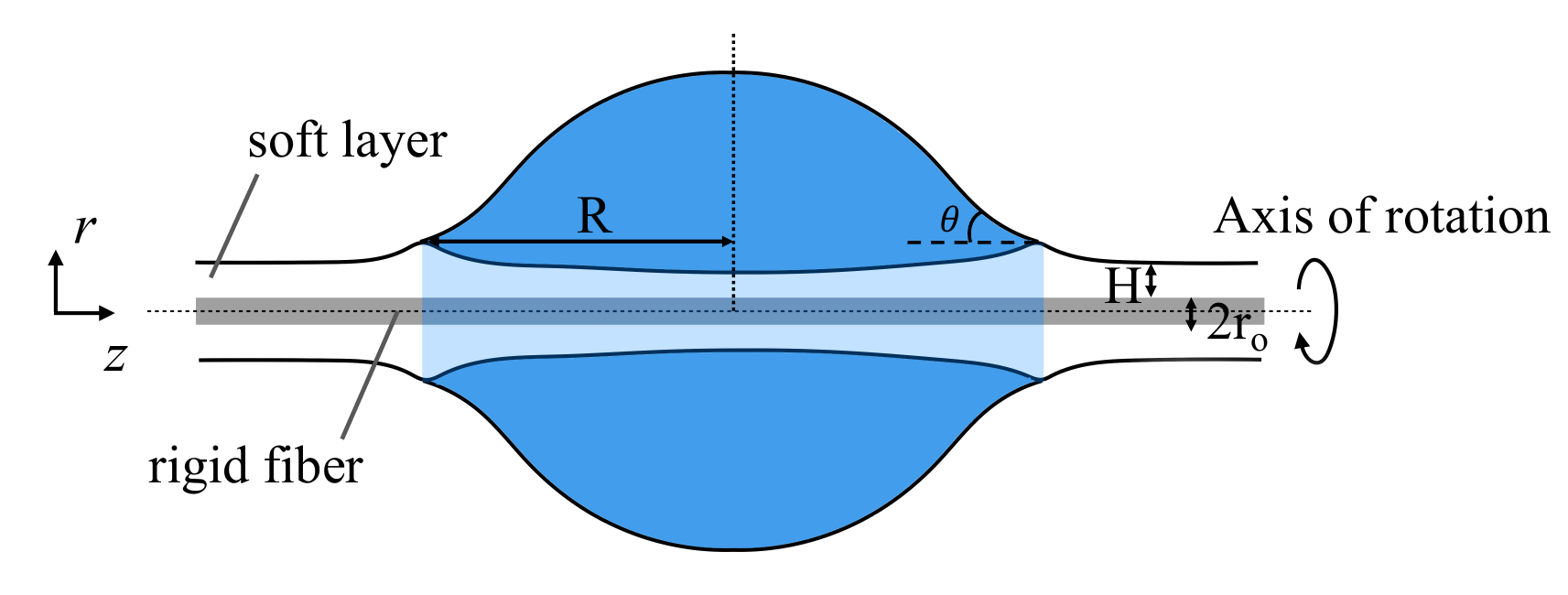}
\caption{Schematic diagram of an axisymmetric barrel-shaped droplet embracing a rigid cylindrical fiber of radius $r_o$ coated with a soft (elastic) solid layer of thickness $H$. The droplet makes an equilibrium contact angle $\theta$ with the fiber. The soft solid layer is deformed by the droplet due to capillary forces.} \label{fig:setup}
\end{center}
\end{figure}

Unlike droplets on planar surfaces,  which appear as a spherical-capped shape given that the Laplace pressure is the dominant force (e.g. effects of gravity are neglected),   the shape of a droplet on a fiber is far more complex.   Extensive studies in the literature \citep{Chou2011,Xu2022} have shown that stable droplets can appear as axisymmetric barrel-shaped or non-axisymmetric clam-shell shaped.  For small equlibrium contact angles  $\theta$ or large droplet sizes relative to the fiber radius,  axisymmetric barrel-shaped droplets tend to be more stable.  When the droplet size ($V^{1/3}$, here $V$ is the droplet volume) is smaller than the fiber radius,  barrel-shaped droplets exist only for $\theta \lesssim 10^{\circ}$ \cite{Xu2022}.      

In this study, we investigate the axisymmetric deformation of a soft solid layer coated on a rigid cylindrical fiber when a  barrel-shaped droplet is embracing it. Previous studies of elastic deformation by wetting often assume a specific situation that $\theta=90^{\circ}$,  and thus the solid surface tension of the soft-solid/liquid interface $\gamma_{sg}$ is the same as that of the soft-solid/gas interface $\gamma_{sl}$ \cite{Style2012,Lubbers2014}.  For stable barrel-shaped droplets on a fiber, the contact angle have to be smaller than $90^{\circ}$. We hence consider generic situations that the two solid surface tensions can be different, i.e. $\gamma_{sg}\neq\gamma_{sl}$. Moreover, it is known that the solid surface tensions not only determine the contact angle at the contact line, but also play a crucial role in suppressing the deformation of a planar soft solid layer  \cite{Jerison2011,Style2012}. How the solid surface tensions affect the soft solid deformation in fiber geometries is studied in details below. 

\section{Formulation}

A schematic of a droplet of volume $V$ resting on a soft-layer-coated fiber is shown in Fig. \ref{fig:setup}.  We consider the droplet Bond number $Bo\equiv \rho g V^{2/3}/\gamma\ll 1$, where $\rho$ is the liquid density, $g$ is the gravitational acceleration and $\gamma$ is the liquid-air surface tension. Hence the effects of gravity on the droplet shape is negligible.  The rigid fiber has a radius $r_o$.  A  soft solid layer of uniform thickness $H$ at an undeformed state is coated on the surface of the rigid fiber.  Due to the axisymmetry of the problem,  we will use the cylindrical coordinate system $(r, \phi,z)$ and the corresponding unit vectors are denoted as ($\hat{\bm{r}},\hat{\bm{\phi}},\hat{\bm{z}}$). The center of mass of the droplet is at the origin, i.e. $r=0$ and $z=0$.  The soft layer is deformed by the droplet. The displacement of the material of the soft layer is denoted as $\bm{U}(r,z)=U_r\hat{\bm{r}}+U_{\phi}\hat{\bm{\phi}}+ U_z\hat{\bm{z}}$.  Due to symmetry, the displacement has a property that $\bm{U}(r,z)=\bm{U}(r,-z)$.  Hence we consider only $z\geq 0$ in the following.  The equilibrium contact angle $\theta$ is defined as the angle of the droplet liquid-air interface at the contact line position $z=R$ with respect to the z-axis as shown in Fig. \ref{fig:setup}.  In this study, we assume that $\theta$ is determined by Young's law \cite{Lubbers2014},  which reads
\begin{eqnarray} \label{young}
\gamma\cos\theta=\gamma_{sg}-\gamma_{sl}.
\end{eqnarray} 
Denoting the elastic stress tensor as $\bm{\sigma}$,  the deformation of the soft layer is governed by the force balance equation
\begin{eqnarray} \label{forcebal}
\nabla\cdot \bm{\sigma}=0.
\end{eqnarray}
As we only consider small slopes of the deformed soft-solid/fluid interface, we employ the linear elastic constitutive model  for the relation between the stress tensor and the displacement, for which the tensor components read
\begin{eqnarray} \label{elst1}
\sigma_{rr} = -p+\frac{E}{(1+\nu)}  \left(\frac{{\partial}U_{r}}{{\partial}r} -\frac{1}{3}\nabla\cdot\bm{U}\right),
\end{eqnarray}

\begin{eqnarray} \label{elst2}
\sigma_{zz} = -p+\frac{E}{(1+\nu)}  \left(\frac{{\partial}U_{z}}{{\partial}z} -\frac{1}{3}\nabla\cdot\bm{U}\right),
\end{eqnarray}

\begin{eqnarray} \label{elst3}
\sigma_{\phi\phi} = -p+\frac{E}{(1+\nu)}  \left(\frac{U_{r}}{r} -\frac{1}{3}\nabla\cdot\bm{U}\right),
\end{eqnarray}

\begin{eqnarray} \label{elst4}\sigma_{rz} = \frac{E}{2(1+\nu)}  \left(\frac{{\partial}U_{r}}{{\partial}z} +\frac{{\partial}U_{z}}{{\partial}r}\right),
\end{eqnarray} 
where the isotropic part of the stress tensor (or the pressure)
\begin{eqnarray} \label{elst5}
p= -\frac{E}{3(1-2\nu)} \nabla\cdot\bm{U},
\end{eqnarray}
 $E$ is the Young modulus and $\nu$ is the Poisson ratio.  Note that $U_{\phi}=0$ due to axisymmetry.

The boundary conditions for the governing equation (\ref{forcebal}) are as follows. Consider that the length of the fiber $2L$ is much larger than the droplet radius $R$, we impose the condition 
\begin{eqnarray} \label{bcdim1}
\bm{U}(r, z=L)=0. 
\end{eqnarray}
 Due to symmetry, at $z=0$,  we have 
 \begin{eqnarray} \label{bcdim2}
U_z(r, z=0)=0
\end{eqnarray}
and
\begin{eqnarray} \label{bcdim3}
\frac{\partial U_r(r, z=0)}{\partial z}=0.
\end{eqnarray}
  At the interface where the soft layer is in contact with the rigid solid, i.e. $r=r_o$, the soft material is undeformed, so we have 
 \begin{eqnarray} \label{bcdim4}
\bm{U}(r=r_o, z)=0.
\end{eqnarray}

For the soft-solid/fluid boundary, i.e. $r=r_o+H$,  we impose a force balance condition.  Here we introduce all the tractions (force per unit area) acting on the interface as the following.  At the contact line, the localized capillary traction $\bm{f}^{l}=\gamma\delta(R-z)(\sin\theta\hat{\bm{r}}-\cos\theta\hat{\bm{z}})$ is pulling the soft layer \cite{Das2011,Bostwick2014}, where $\delta(z)$ is the Dirac delta function. Second, the Laplace pressure inside the droplet can also deform the soft layer.  The traction $\bm{f}^{La}$ due to the Laplace pressure is given as $\bm{f}^{La}=-\gamma \kappa_l H_s(R- z )\hat{\bm{r}}$,  where $\kappa_l$ is the curvature of the droplet's liquid-air interface and $H_s(z )$ is Heaviside step function.   Third,  the elastic traction due to soft solid deformation is $\bm{f}^{el}=-(\sigma_{rr}\hat{\bm{r}}+\sigma_{rz}\hat{\bm{z}})\vert_{r=r_o+H}$.   Fourth, the contribution from the soft solid surface tension gives a traction $\bm{f}^{s}=-\gamma_{s}\kappa_{s}\hat{\bm{r}}+\partial{\gamma_s}/\partial{z}\hat{\bm{z}}$, where $\gamma_{s}$ is the soft solid surface tension and $\kappa_{s}$ is the curvature of the deformed soft-solid/fluid interface. We assume that the solid surface stress is the same as the solid surface energy, and call them the solid surface tension. In other words, we neglect the Shutterworth effect \citep{Shuttleworth1950,Andreotti2016,Masurel2019}. Note that the assumption of a small soft-solid/fluid interface slope  has been used to obtain the expression of $\bm{f}^{s}$. For small interface slopes, the curvature can be simplified as $\kappa_{s}=\left[1/(r_o+H+U_r)- {\partial^2}U_r/{\partial}z^2\right]\vert_{r=r_o+H}$. 
 For the soft solid surface tension, there is a jump across the contact line, meaning that $\gamma_{s}=\gamma_{sl}+(\gamma_{sg}-\gamma_{sl})H_s(R-z )$. Thus, $\bm{f}^{s}=-\gamma_{s}\kappa_{s}\hat{\bm{r}}+(\gamma_{sg}-\gamma_{sl})\delta(R-z)\hat{\bm{z}}$.  Balancing all the tractions we have the following boundary condition at $r=r_o+H$,
\begin{eqnarray} \label{forceBL}
\bm{f}^{l}+\bm{f}^{La}+\bm{f}^{el}+\bm{f}^{s}=0.
\end{eqnarray} 

Regarding the shape of the droplet's liquid-air interface, described by $r=g(z)$, the profile is obtained by solving the equation of uniform Laplace pressure, which implies $\kappa_{l}$ has a constant value, i.e. $\kappa_{l}=k_c$ . We describe the details of the governing equation and the corresponding boundary conditions at the later part of this section.

Next, we non-dimensionalize the variables as the following.  We rescale the coordinates and displacements by $H$ and the pressure by $E$, namely
\begin{eqnarray}
\tilde{r}=\frac{r}{H}, \
\tilde{z}=\frac{z}{H}, \
\tilde{U}_r=\frac{U_r}{H}, \
\tilde{U}_z=\frac{U_z}{H}, \ 
\tilde{p}=\frac{3p}{E}.
\end{eqnarray}

For the elastic deformation,  the dimensionless form of the governing equation (\ref{forcebal}) written in vector components  are
\begin{eqnarray} \label{ges1}
\tilde{\nabla}^{2}\tilde{U}_r - \frac{\tilde{U}_r}{\tilde{r}^{2}} -\frac{\partial \tilde{p}}{{\partial}\tilde{r}}=0
\end{eqnarray} 
in r-direction and
\begin{eqnarray} \label{ges2}
\tilde{\nabla}^{2}\tilde{U}_z - \frac{\partial \tilde{p}}{{\partial}\tilde{z}}=0
\end{eqnarray} 
in z-direction. The dimensionless form of the boundary conditions (\ref{bcdim1})-(\ref{bcdim4}) are
\begin{eqnarray} \label{bcs1}
\tilde{\bm{U}}(\tilde{r}, \tilde{z}= \tilde{L})=0,
\end{eqnarray}  
\begin{eqnarray} \label{bcs2}
\tilde{U_z}(\tilde{r}, \tilde{z}=0)=0,
\end{eqnarray} 
\begin{eqnarray} \label{bcs3}
\frac{\partial \tilde{U_r}}{\partial \tilde{z}}(\tilde{r}, \tilde{z}=0)=0,
\end{eqnarray} 
and
\begin{eqnarray} \label{bcs4}
\tilde{\bm{U}}(\tilde{r}=\tilde{r}_o, \tilde{z})=0,
\end{eqnarray}
respectively, where $\tilde{r}_o=r_o/H$ and $\tilde{L}=L/H$.

For the soft-solid/fluid interface, we first define $\tilde{u}_r\equiv \tilde{U}_r(\tilde{r}=\tilde{r}_o+\tilde{H}, \tilde{z})$ and  $\tilde{u}_z\equiv \tilde{U}_z(\tilde{r}=\tilde{r}_o+\tilde{H}, \tilde{z})$. The dimensionless form of the force balance condition (\ref{forceBL}), written in vector components, are given as follows. For the r-components, we obtain
%\begin{eqnarray} \label{bcs5}
%\sin\theta\delta (\tilde{R}-\tilde{z})-\tilde{k}_cH(\tilde{R}-\tilde{z}) \nonumber\\
% +\frac{\tilde{E}}{3}\left(\tilde{p} -\frac{3}{1+\nu} \frac{{\partial}\tilde{u}_{r}}{{\partial}\tilde{r}}\right)\nonumber \\
%+\left[\tilde{\gamma}_{sl}+(\tilde{\gamma}_{sg}-\tilde{\gamma}_{sl})H_s(\tilde{R}-\tilde{z} )\right]\left[ \frac{1}{\tilde{r}_o+1+ \tilde{u}_r}- \frac{{\partial}^2\tilde{u}_{r}}{{\partial}{\tilde{r}}^2} \right]\nonumber\\
%=0,
%\end{eqnarray}
\begin{equation} \label{bcs5}
\begin{split}
\sin\theta\delta (\tilde{R}-\tilde{z})-\tilde{k}_cH(\tilde{R}-\tilde{z}) \\
 +\frac{\tilde{E}}{3}\left(\tilde{p} -\frac{3}{1+\nu} \frac{{\partial}\tilde{u}_{r}}{{\partial}\tilde{r}}\right) \\
+\left[\tilde{\gamma}_{sl}+(\tilde{\gamma}_{sg}-\tilde{\gamma}_{sl})H_s(\tilde{R}-\tilde{z} )\right]\left[ \frac{1}{\tilde{r}_o+1+ \tilde{u}_r}- \frac{{\partial}^2\tilde{u}_{r}}{{\partial}{\tilde{r}}^2} \right] \\
=0,
\end{split}
\end{equation}
where the dimensionless parameters are defined as  
\begin{eqnarray}
\tilde{E}\equiv\frac{EH}{\gamma}, \
\tilde{\gamma}_{sl}\equiv\frac{\gamma_{sl}}{\gamma}, \
\tilde{\gamma}_{sg}\equiv\frac{\gamma_{sg}}{\gamma}, \
\tilde{k}_c\equiv k_cH, \
\tilde{R}\equiv\frac{R}{H}.
\end{eqnarray}
For the z-components of the force balance (\ref{forceBL}), after using Young's law (\ref{young}), $-\gamma \cos\theta \delta(R-z)$ cancels out $(\gamma_{sg}-\gamma_{sl})\delta(R-z)$, hence we obtain the vanishing elastic shear stress condition
\begin{eqnarray} \label{bcs6}
\frac{{\partial}\tilde{u}_{r}}{{\partial}\tilde{z}} +\frac{{\partial}\tilde{u}_{z}}{{\partial}\tilde{r}}=0.
\end{eqnarray}

For the droplet profile,  the dimensionless form of the uniform curvature equation is
\begin{eqnarray} \label{droppro}
\tilde{\kappa}_l\equiv \frac{1}{\tilde{g}\sqrt{1+(\frac{\partial{\tilde{g}}}{\partial{\tilde{z}}})^2}}-\frac{\frac{\partial^2{\tilde{g}}}{\partial{\tilde{z}^2}}}{\sqrt{1+(\frac{\partial{\tilde{g}}}{\partial{\tilde{z}}})^2}^3}=\tilde{k}_c,
\end{eqnarray}
where $\tilde{g}=g/H$.
The boundary conditions at the contact line are: 
\begin{eqnarray}\label{dropbc1}
\tilde{g}( \tilde{z}=\tilde{R})=\tilde{r}_o+1+\tilde{u}_r(\tilde{r}=\tilde{r}_o+1,\tilde{z}=\tilde{R})
\end{eqnarray}
and
\begin{eqnarray} \label{dropbc2}
\frac{\partial{\tilde{g}( \tilde{z}=\tilde{R})}}{\partial{\tilde{z}}}=-\tan\theta.
\end{eqnarray}
The droplet volume is given by
 \begin{eqnarray} \label{dropvol}
\tilde{V}\equiv\frac{V}{H^3}=2\pi\int^{\tilde{R}}_{0}\tilde{g}^2-(\tilde{r}_o+1+\tilde{u}_r)^2\operatorname{d}{\tilde{z}}.
\end{eqnarray}

In this study, we consider the soft material to be incompressible, which means $\nu=0.5$. The incompressibility condition implies
\begin{eqnarray} \label{contin}
\tilde{\nabla}\cdot \tilde{\bm{U}}=0.
\end{eqnarray}
The governing equations (\ref{ges1}), (\ref{ges2}), (\ref{droppro}) and (\ref{contin}) are solved together with the conditions  (\ref{bcs1})- (\ref{bcs5}), (\ref{bcs6}), (\ref{dropbc1})- (\ref{dropvol}) using the finite element method for which the details are given in the Appendix. The dimensionless control parameters are $\tilde{r}_o$, $\tilde{E}$, $\tilde{\gamma}_{sl}$, $\tilde{\gamma}_{sg}$ and $\tilde{V}$.

\section{Results}
\subsection{Large droplet limit}\label{lardro}
First we look at situations in which the droplet length scale ($V^{1/3}$) is much larger than the other length scales: $H$, $r_o$,  $\gamma/E$, $\gamma_{sl}/E$ and $\gamma_{sg}/E$. 
The Laplace pressure $\bm{f}^{La}$ is neglected in this limit, and thus the only external capillary force acting on the soft solid is the localized force $\bm{f}^{l}$ at the contact line. We will demonstrate how the solid deformation varies with the change of $\tilde{r}_o$ and $\tilde{E}$ for both cases of $\gamma_{sg}=\gamma_{sl}$ (i.e. $\theta=90^{\circ}$) and $\gamma_{sg}\neq\gamma_{sl}$ (i.e. $\theta\neq 90^{\circ}$).    

\subsubsection{\textbf{Cases of $\gamma_{sg}=\gamma_{sl}$ (i.e. $\theta=90^{\circ}$)}} \label{sym}
\begin{figure}
\begin{center}
\includegraphics[width=0.45\textwidth]{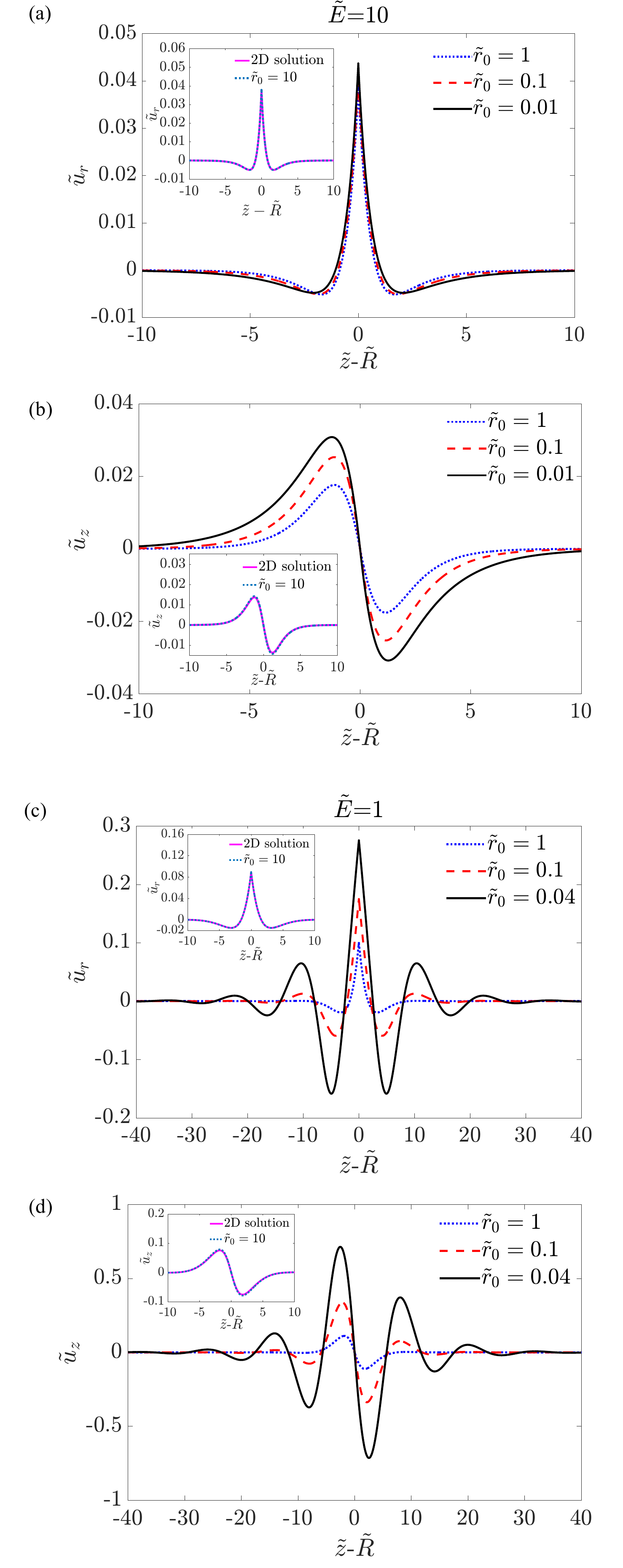}
\caption{The rescaled displacements $\tilde{u}_r$ and $\tilde{u}_z$ as a function of $\tilde{z}-\tilde{R}$ for different values of $\tilde{r}_0$ and $\tilde{E}=10$ in (a)-(b), and $\tilde{E}=1$ in (c)-(d). Insets: the analytical solutions of 2D plain strain case (solid lines)\cite{Style2012} and our numerical results for $\tilde{r}_0=10$ (dotted lines).  Other parameters: $\tilde{\gamma}_{sl}=\tilde{\gamma}_{sg}=5$.} \label{fig2}
\end{center}
\end{figure}

 We look at two different cases of softness, namely $\tilde{E}=1$ and $\tilde{E}=10$.   In Fig.  \ref{fig2},  the displacements at the soft-solid/fluid interface $\tilde{u}_r$ and $\tilde{u}_z$ are plotted as a function of $ \tilde{z}-\tilde{R} $ for different values of   $\tilde{r}_o$. To validate our computations,  we first compare our results for a large fiber radius $\tilde{r}_o=10$ with the  analytical solution of the 2D plain strain case \cite{Style2012}. In the insets of Fig.  \ref{fig2}, we show that our numerical results collapse with the analytical 2D solutions.  Next we look at how the fiber radius modifies the deformation.   When the layer is stiff ($\tilde{E}=10$), as shown in Fig. \ref{fig2} (a),  the out-of-plane displacement $\tilde{u}_r$ is insensitive to the value of $\tilde{r}_o$. Relatively, there is a stronger dependence of the in-plane displacement $\tilde{u}_z$ on $\tilde{r}_o$ as shown in Fig.  \ref{fig2} (b). In contrast, for the softer case of $\tilde{E}=1$ shown in Fig.  \ref{fig2} (c) and (d), the magnitude of displacements increase significantly with decreasing $\tilde{r}_o$ when  $\tilde{r}_o \leq 1$.   Moreover, unlike the stiff or large rigid fiber radius cases, in which the magnitude of displacement decays quickly with distances from the contact line, the oscillations are significant for the case of $\tilde{E}=1$ when $\tilde{r}_o \ll 1$.  

How do we understand the results? For $\tilde{r}_o \gg 1$, the deformation behaves the same as the 2D solution as one might  expect.  The more interesting regime is when $\tilde{r}_o \lesssim 1$.  When the material is stiff, namely $\tilde{E}\gg1$ (i.e. $\gamma/E\ll H$), the maximum out-of-plane displacement $\tilde{u}_r$ at the contact line is  independent of $\tilde{r}_o$.   This can be understood due to the fact that  the deformation is small compared to the thickness of the soft layer.  The displacements vanish at a short distance from the soft-solid/fluid interface. Hence the no-displacement condition at $\tilde{r}=\tilde{r}_o$ only plays as a small correction to the displacements. 
On the other hand, when $\tilde{E}\lesssim 1$, the bulk material deforms more significantly, hence the boundary conditions at $\tilde{r}_o$ have a stronger effect. Furthermore, not only the bulk elastic stresses, but the azimuthal curvature of the soft solid interface, i.e. $1/(\tilde{r}_o+1+\tilde{u}_r)$,  also plays a crucial role for the strong displacement oscillations. We have checked that removing  the azimuthal curvature term in Eq. (\ref{bcs5}) results in the disappearance of displacement oscillations.
 
\subsubsection{Cases of $\gamma_{sg}\neq\gamma_{sl}$ (i.e. $\theta\neq 90^{\circ}$)}\label{asym}

\begin{figure}
\begin{center}
\includegraphics[width=0.45\textwidth]{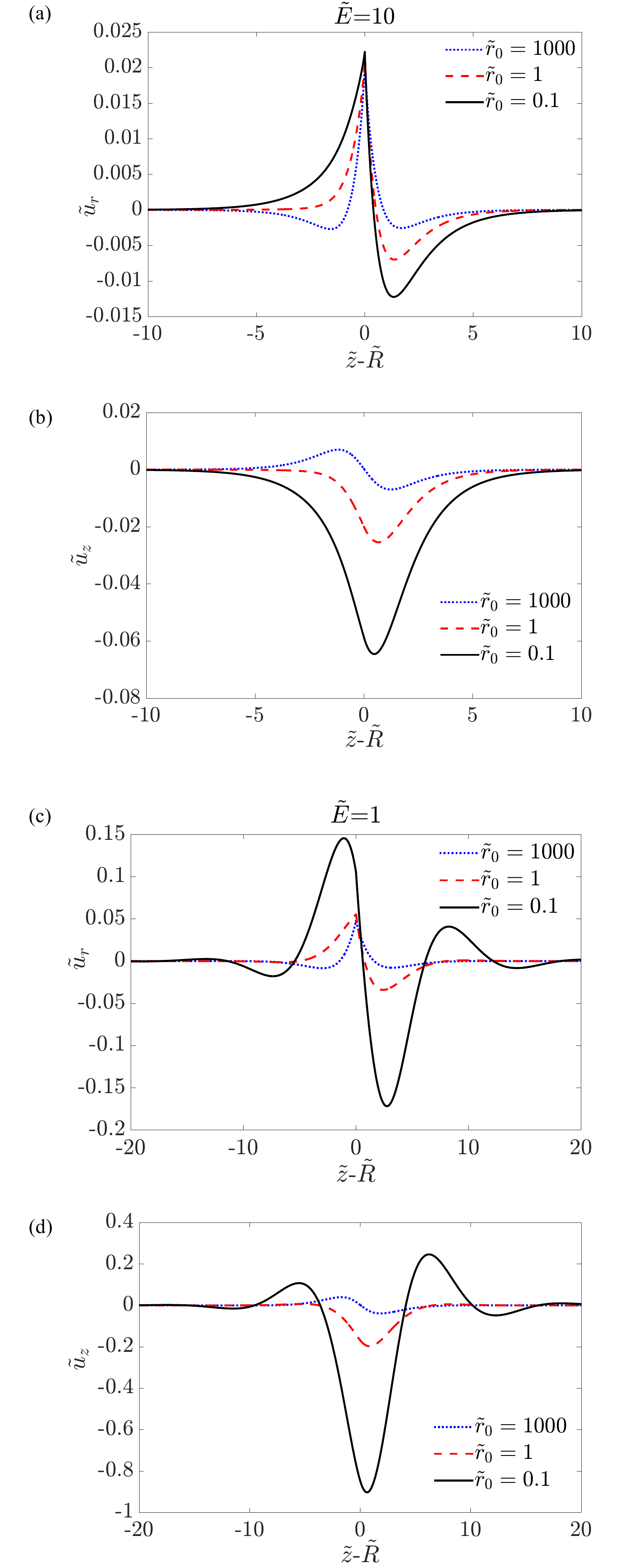}
\caption{The rescaled displacements $\tilde{u}_r$ and $\tilde{u}_z$ as a function of $\tilde{z}-\tilde{R}$ for different values of $\tilde{r}_0$ and $\tilde{E}=10$ in (a)-(b), and $\tilde{E}=1$ in (c)-(d). Other parameters: $\tilde{\gamma}_{sl}=4.13$ and $\tilde{\gamma}_{sg}=5$.} \label{fig3}
\end{center}
\end{figure}
 We take  $\tilde{\gamma}_{sl}=4.13$, $\tilde{\gamma}_{sg}=5$, and thus $\theta=30^{\circ}$ using Young's law (\ref{young}). In Fig.  \ref{fig3}, we plot the displacements  $\tilde{u}_r$ and $\tilde{u}_z$ as a function of $\tilde{z}-\tilde{R} $. For $\tilde{E}=10$, although the maximum of $\tilde{u}_r$ is almost independent of $\tilde{r}_o$, the dimple (minimum) depends significantly on $\tilde{r}_o$ as shown in Fig. \ref{fig3} (a). When the rigid fiber radius is large, for example, for $\tilde{r}_o=1000$, the dimple on the solid-liquid side (left) is slightly larger than the solid-gas side (right). This is consistent with the finding of previous studies for plate cases that the solid surface tension suppresses deformation \cite{Jerison2011}. Note that $\tilde{\gamma}_{sl}<\tilde{\gamma}_{sg}$. However, when decreasing $\tilde{r}_o$, $\tilde{u}_r$ increases on the solid/liquid side and decreases on the solid/gas side. Hence, the dimple of the solid/liquid interface decreases and disappears. On the other hand, the dimple of the solid/gas interface becomes larger. This disparity can be observed clearly for $\tilde{r}_o=0.1$ in Fig. \ref{fig3} (a). 
 
% Another remarkable feature is about the interface slope at the contact line, i.e. at $\tilde{z}=\tilde{R}$. It is known that the interface slope is discontinuous when crossing from the solid-liquid interface to the solid-air interface, for which the values are determined by Neumanns's relation \cite{Andreotti2020}. When decreasing $\tilde{r}_o$, as we can see in Fig. \ref{fig3} (a), the interface slopes at the contact line on both sides decrease. The change of interface slope at the contact line (or named the contact angle rotation) has also been demonstrated in previous studies of a droplet on a soft planar substrate when varying the softness parameter \cite{Style2013,Lubbers2014,Hui2014,Dervaux2015}. 
  
 The in-plane displacement $\tilde{u}_z$ shown in Fig. \ref{fig3} (b) also demonstrates specific features when varying $\tilde{r}_o$. For large $\tilde{r}_o$, the material around the contact line at the interface displaces towards the contact line for both sides. When decreasing $\tilde{r}_o$, $\tilde{u}_z$ decreases. For  $\tilde{r}_o\lesssim 1$,  $\tilde{u}_z$ is negative for the whole interface, meaning that all the material at the interface displaces towards the side of interface with a smaller solid surface tension (left). 
 
 For a softer material ($\tilde{E}=1$) shown in Fig. \ref{fig3} (c) and (d), the behavior of $\tilde{u}_r$ and $\tilde{u}_z$ when decreasing $\tilde{r}_o$ is similar to $\tilde{E}=10$. Remarkably when $\tilde{r}_o$ is reduced to $\tilde{r}_o = 0.1$, $\tilde{u}_r$ around the contact line on the solid-liquid side increases to values larger than $\tilde{u}_r$ at the contact line. Hence, the maximum of $\tilde{u}_r$ is not at the contact line position but shifts to the solid/liquid side as shown in Fig. \ref{fig3} (c). Similar to that we have seen for the case of $\theta= 90^{\circ}$, pronounced oscillations of $\tilde{u}_r$ and $\tilde{u}_z$ appear when $\tilde{r}_o \ll 1$.
\subsubsection{Dependence on the contact angle (or the solid surface tension difference) }

\begin{figure}
\begin{center}
\includegraphics[width=0.45\textwidth]{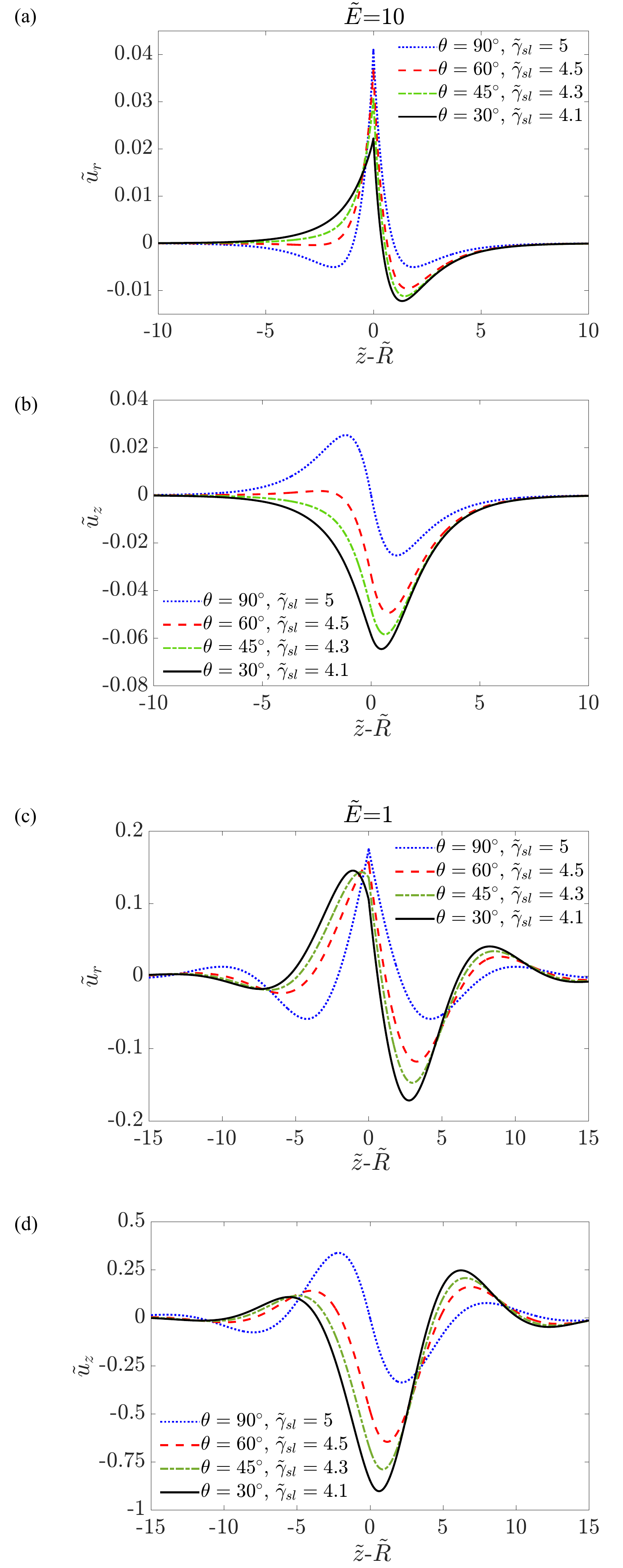}
\caption{The rescaled displacements $\tilde{u}_r$ and $\tilde{u}_z$ as a function of $\tilde{z}-\tilde{R}$ for different values of $\tilde{\gamma}_{sl}$ and $\tilde{E}=10$ in (a)-(b), and $\tilde{E}=1$ in (c)-(d). Other parameters: $\tilde{\gamma}_{sg}=5$ and $\tilde{r}_o=0.1$.} \label{fig4}
\end{center}
\end{figure}

We look at how the features of deformation change when the contact angle is varied from $\theta=90^{\circ}$ to $\theta=30^{\circ}$ for a fixed value of $\tilde{r}_o=0.1$. According to Young's law, changing $\theta$ also means that $\tilde{\gamma}_{sg}-\tilde{\gamma}_{sl}$ is varied. We keep $\tilde{\gamma}_{sg}=5$ for all cases and change the value of $\tilde{\gamma}_{sl}$. We again consider two different values of $\tilde{E}$. The displacements  $\tilde{u}_r$ and $\tilde{u}_z$ as a function of $\tilde{z}-\tilde{R}$ for different $\theta$ (or $\tilde{\gamma}_{sl}$) are plotted in Fig. \ref{fig4}. 

In Fig. \ref{fig4} (a), we can see that the maximum of $\tilde{u}_r$ decreases when reducing $\theta$ due to the fact that the pulling contact line force scales as $\gamma \sin\theta$. As $\theta$ is decreased, the asymmetry of displacements between the solid-liquid side and the solid-gas side becomes more apparent. When comparing with the results in Fig. \ref{fig3}, we obsserve that the enhancement of asymmetry when reducing $\theta$ (for a fixed value of $\tilde{r}_o$) is similar to that of decreasing $\tilde{r}_o$ (for a fixed value of $\theta$). In the following we point out some of these similar trends. As shown in Fig. \ref{fig3} (a) or in Fig. \ref{fig4} (a) for $\tilde{E}=10$, when reducing $\tilde{r}_o$ or $\theta$, the dimple of $\tilde{u}_r$ on the solid-liquid side  (with a smaller solid surface tension) decreases and the dimple on the solid-gas side becomes larger. For $\tilde{E}=1$, the maximum of $\tilde{u}_r$ shifts from the contact line position to the solid-liquid side when $\tilde{r}_o$ or $\theta$ is reduced to below a certain value as shown in Fig. \ref{fig3} (c) or in Fig. \ref{fig4}(c). For the in-plane displacement shown in Fig. \ref{fig3} (b) and Fig. \ref{fig3} (d), or Fig. \ref{fig4} (b) and Fig. \ref{fig4} (d), $\tilde{u}_z$ around the contact line decreases when reducing $\tilde{r}_o$ or $\theta$. This means that the material at the interface displaces more to the side of interface with a smaller solid surface tension (left).

\subsection{Dependence on the droplet size}
\begin{figure}
\begin{center}
\includegraphics[width=0.45\textwidth]{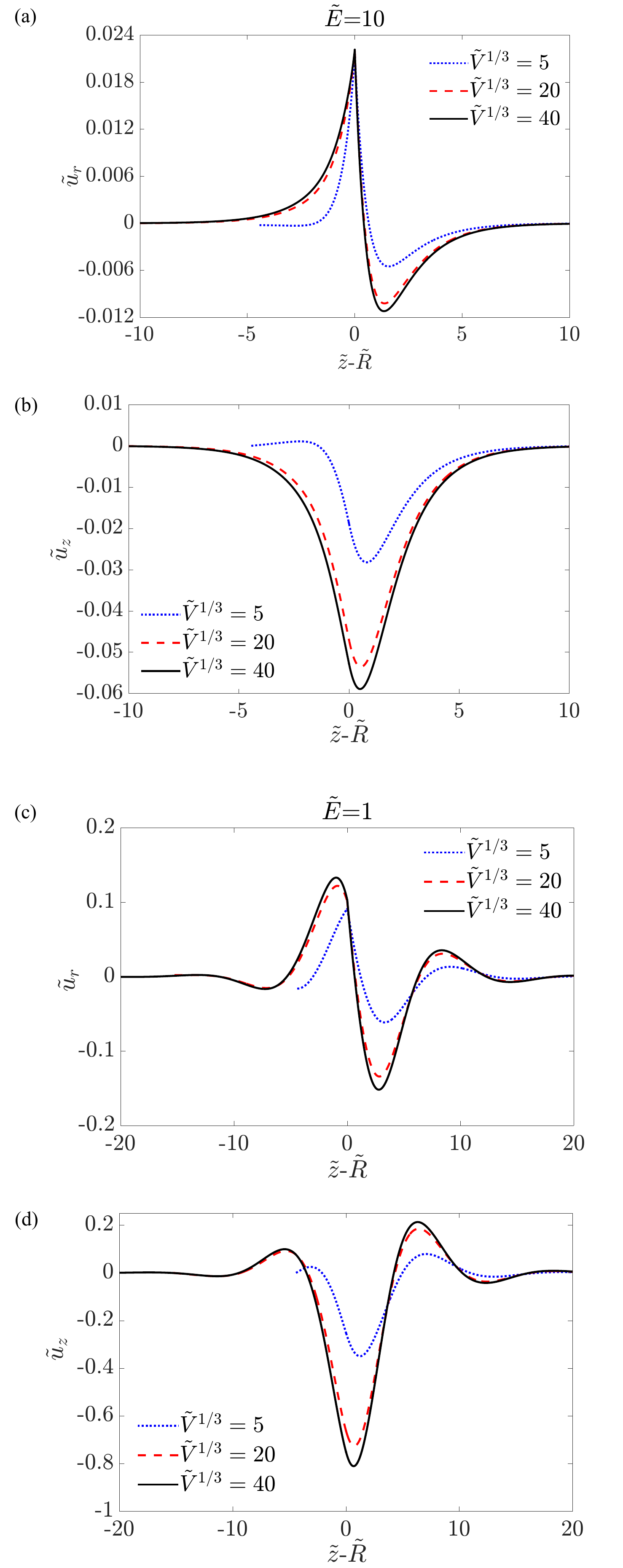}
\caption{The rescaled displacements $\tilde{u}_r$ and $\tilde{u}_z$ as a function of $\tilde{z}-\tilde{R}$ for different values of $\tilde{V}^{1/3}$ and $\tilde{E}=10$ in (a)-(b), and $\tilde{E}=1$ in (c)-(d). Other parameters: $\tilde{\gamma}_{sl}=4.134$, $\tilde{\gamma}_{sg}=5$ and $\tilde{r}_o=0.1$.} \label{fig5}
\end{center}
\end{figure}

For a droplet of finite size, the traction term due to the Laplace pressure $\bm{f}^{La}$ is included in our computations.  We choose  $\theta=30^{\circ}$ and $\tilde{r}_o=0.1$. Remind that to obtain a stable axisymmetric barrel-shaped droplet, the smallest droplet size has to be comparable to the total fiber radius. Hence we consider 3 different droplet sizes: $\tilde{V}^{1/3}=5$, 20 and 40. In Fig. \ref{fig5}, the displacements $\tilde{u}_r$ and $\tilde{u}_z$ are plotted as a function of $\tilde{z}-\tilde{R} $. We can see that the curves for $\tilde{V}^{1/3}=20$ and 40 are close to results for the large droplet limit. For $\tilde{V}^{1/3}=5$, the Laplace pressure term $\bm{f}^{La}$ tends to reduce the magnitudes of displacements and oscillations, except $\tilde{u}_r$ at the contact line which remains the same.

\section{Conclusion}

We study the elastic deformation of a soft layer coated on a rigid cylindrical fiber when an axisymmetric barrel-shaped droplet is embracing it. For a droplet contact angle $\theta=90^{\circ}$, and thus $\gamma_{sl}=\gamma_{sg}$ according to Young's law, we find that the magnitudes of both displacements  $\tilde{u}_r$ and $\tilde{u}_z$ increase with decreasing $\tilde{r}_0$. For $\theta\neq 90^{\circ}$  (i.e. $\gamma_{sl}\neq\gamma_{sg}$), the deformations on the solid-liquid side and  the solid-gas side are different. This disparity of deformation is enhanced when decreasing    $\tilde{r}_0$ (for a fixed value of $\theta$), or decreasing $\theta$ (for a fixed value of $\tilde{r}_0$). The dimple of $\tilde{u}_r$ on the side with a smaller solid surface tension becomes smaller while the dimple becomes larger on the other side. 

Pronounced oscillations of displacements are observed for the cases of $\tilde{E}<1$ and $\tilde{r}_o \lesssim 1$. This slow decay of deformation with distances from the contact line position suggests a relatively long-range interaction between droplets on a soft-layer-coated fiber. Hence, it is expected that droplet migration and interaction are significantly different from those observed on a planar soft substrate or a purely rigid fiber, for example, the inverted Cheerios effect \cite{Karpitschka2016} and coalescence \cite{Lee2022}, which remain open questions to be explored.

\section{Appendix: Finite element method}

The non-dimensional deformation of the soft layer of the fiber and the droplet profile are computed by solving the governing equations  
 (\ref{ges1}), (\ref{ges2}), (\ref{droppro}) and (\ref{contin}) together with the conditions  (\ref{bcs1})- (\ref{bcs5}), (\ref{bcs6}), (\ref{dropbc1})- (\ref{dropvol}) by using a finite element method (FEM). We discretize our non-dimensional variables with linear elements and solve the coupled equations using a Newton solver from the FEM library FEniCS \cite{logg2012automated}.

 For the Dirac delta function, we approximate it with a Gaussian function as $\delta (R-z) \approx W (R-z)=exp[-(R-z)^2/(2l_{m}^2)]/(l_{m}\sqrt{2\pi})$, where $l_m$ can be interpreted as a microscopic length (e.g. interface thickness) such that $l_m \ll H $. In the limit that $l_m \rightarrow 0 $, $\delta (R-z) = W (R-z)$. In this study, $l_m/H = 10^{-5}$ is set for all the cases. We have used the adaptive mesh sizing such that the smallest mesh size near the contact line is $dx_1/H=10^{-7}$ and the largest mesh size far away from the contact line is $dx_0/H=0.05$.

The mesh convergence of the numerical solver is tested with the case of the large droplet limit and the condition of $\tilde{r}_0 \gg 1$. In this limit, the deformation of the soft layer will converge towards the 2D soft plate case. In Figure 6 we compare our numerical results with the analytical solution for the 2D soft plate case by plotting the deformation of the soft solid for three different mesh resolutions namely $dx_0/H=0.1$, $0.05$ and $0.025$. For all three mesh sizes, the numerical solutions are in a good agreement with the analytical solution.

\begin{figure}
\begin{center}
\includegraphics[width=0.45\textwidth]{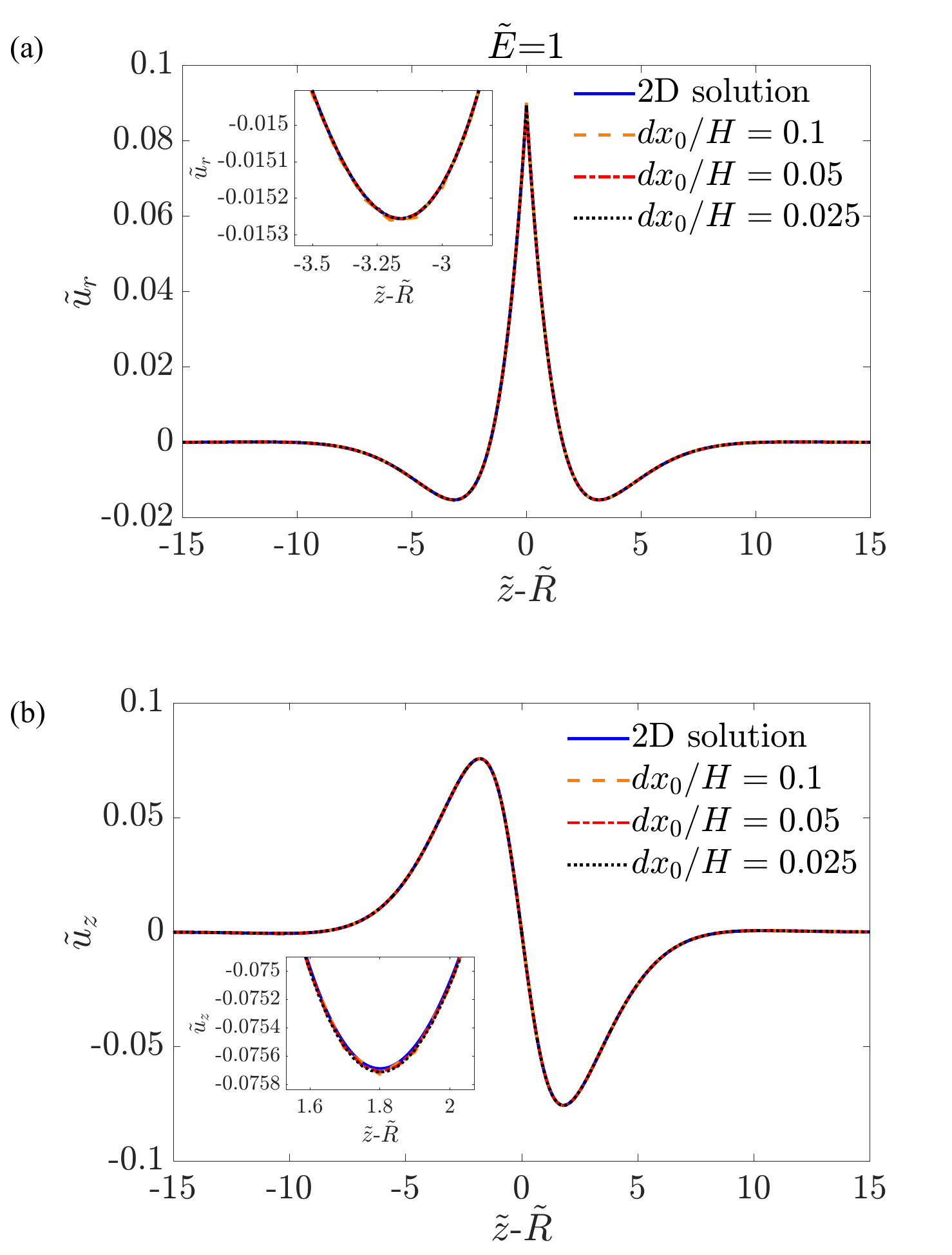}
\caption{Convergence of the mesh. The rescaled displacements $\tilde{u}_r$ and $\tilde{u}_z$ as a function of $\tilde{z}-\tilde{R}$ for different mesh sizes. Other parameters: $\theta=90^{\circ}$, $\tilde{\gamma}_{sl}=\tilde{\gamma}_{sg}=5$, $\tilde{r}_o=1000$.} \label{fig6}
\end{center}
\end{figure}

\section*{Conflicts of interest}
``There are no conflicts to declare''.

\section*{Acknowledgements}
The authors gratefully acknowledges financial support from the Research Council of Norway (Project No. 315110). BXZ thanks Stephane Poulain, Jarle Sogn and Miroslav Kuchta for the discussions of the numerical method.

\bibliography{DropOnSoftFiber}

\end{document}